\documentclass[iop,apjl,tighten]{emulateapj}

\usepackage{natbib}
\usepackage{graphicx}
\usepackage{hyperref}

\shorttitle{XO-2 ELEMENTAL ABUNDANCES}
\shortauthors{RAM\'IREZ ET AL.}
\submitted{To appear in the Astrophysical Journal}

\newcommand{\feh}{\mathrm{[Fe/H]}}
\newcommand{\teff}{T_\mathrm{eff}}
\newcommand{\logg}{\log g}
\newcommand{\vt}{v_t}

\newcommand{\fei}{\ion{Fe}{1}}
\newcommand{\feii}{\ion{Fe}{2}}
\newcommand{\kms}{km\,s$^{-1}$}
\newcommand{\tc}{T_\mathrm{C}}

\begin{document}

\title{THE DISSIMILAR CHEMICAL COMPOSITION OF THE PLANET-HOSTING \\
       STARS OF THE XO-2 BINARY SYSTEM}

\author{I.\,Ram\'irez\altaffilmark{1},
        S. Khanal\altaffilmark{1},
        P. Aleo\altaffilmark{1},
        A. Sobotka\altaffilmark{1},
        F.\ Liu,\altaffilmark{2}
        L.\ Casagrande,\altaffilmark{2} \\
        J.\ Mel\'endez,\altaffilmark{3}
        D.\ Yong,\altaffilmark{2}
        D.\,L.\,Lambert,\altaffilmark{1}
        and
        M.\ Asplund\altaffilmark{2}
        }
\altaffiltext{1}{McDonald Observatory and Department of Astronomy,
                 University of Texas at Austin;
                 2515 Speedway, Stop C1402, Austin, TX 78712-1205, USA}
\altaffiltext{2}{Research School of Astronomy and Astrophysics,
                 Australian National University;
                 Mt.\ Stromlo Observatory, via Cotter Rd., Weston, ACT\,2611,
                 Australia}
\altaffiltext{3}{Departamento de Astronomia do IAG/USP,
                 Universidade de S\~ao Paulo;
                 Rua do M\~atao 1226, S\~ao Paulo, 05508-900, SP, Brasil}

\begin{abstract}
Using high-quality spectra of the twin stars in the XO-2 binary system, we have detected significant differences in the chemical composition of their photospheres. The differences correlate strongly with the elements' dust condensation temperature. In XO-2N, volatiles are enhanced by about 0.015\,dex and refractories are overabundant by up to 0.090\,dex. On average, our error bar in relative abundance is 0.012\,dex. We present an early metal-depletion scenario in which the formation of the gas giant planets known to exist around these stars is responsible for a 0.015\,dex offset in the abundances of all elements while $20\,M_\oplus$ of non-detected rocky objects that formed around XO-2S explain the additional refractory-element difference. An alternative explanation involves the late accretion of at least $20\,M_\oplus$ of planet-like material by XO-2N, allegedly as a result of the migration of the hot Jupiter detected around that star. Dust cleansing by a nearby hot star as well as age or Galactic birthplace effects can be ruled out as valid explanations for this phenomenon.
\end{abstract}


\section{INTRODUCTION} \label{s:intro}

Over the past few years our team has pioneered and further developed a technique for measuring elemental abundances in stars at the highest precision possible \cite[e.g.,][]{melendez09:twins,melendez12,ramirez09,ramirez11,ramirez14:bst,ramirez14:harps,bedell14,liu14,tucci-maia14}. While most other chemical analyses are limited by typical error bars of order 0.05\,dex ($\sim$10\,\%) in quantities such as [Fe/H], [O/Fe], etc.,\footnote{We use the standard notation: $A_\mathrm{X}=\log(n_\mathrm{X}/n_\mathrm{H})+12$, where $n_\mathrm{X}$ is the number density of element X, $\mathrm{[X/H]}=A_\mathrm{X}-A_\mathrm{X}^\odot$, and $\mathrm{[X/X']}=\mathrm{[X/H]}-\mathrm{[X'/H]}$.} we have been able to derive relative abundances with 0.01\,dex ($\sim$2\,\%) precision, and even better, down to the 1\,\% level, in a few critical cases.

The success of our technique can be traced back to two key requirements: 1) a careful selection of stars which are very similar to each other (stellar twins), and 2) a strict differential analysis using spectra of extremely high quality. When studying samples of stellar twins, the systematic uncertainties that plague chemical abundance analysis are so similar among all the sample stars that a strict differential analysis (e.g., measurement of relative abundances on a line-by-line basis) essentially cancel out. This leaves the observational noise as the main source of error. Thus, error bars can be made very small simply by acquiring high-resolution spectra of very high signal-to-noise ratio ($S/N$). Instead of the typical $S/N\sim100$ (or somewhat higher when given per resolution element) used in standard chemical analyses, all our high-precision abundance studies have employed spectra of $S/N\gtrsim300$ {\it per pixel} (and typically higher than 400) and resolution $R=\lambda/\Delta\lambda\gtrsim60\,000$.

Thanks to the high precision of our derived abundances, we have uncovered interesting abundance anomalies among stars known to host planets of different kinds. \cite{melendez09:twins} first showed that the solar photosphere is slightly deficient in refractory elements relative to volatiles when compared to a sample of solar twin stars \cite[see also][]{ramirez09,ramirez10}. They attributed this deficiency to the formation of rocky material in the solar system. Essentially, they claim that the missing mass of refractories in the Sun was locked-up in the terrestrial planets, asteroids, and other solar system rocks at the time of star and planet formation.

Later, \cite{ramirez11} showed that the secondary star of the 16\,Cygni binary system, which is known to host a gas giant planet, is slightly metal-poor relative to the primary, which has not yet shown evidence of hosting planets. Under the assumption that stars in multiple systems must have had the same initial chemical composition, this observation was interpreted as a signature of the formation of the gas giant planet, which took {\it both} refractory and volatile elements from its host star during the star/planet formation stage.

Since the bulk metallicity of gas giant planets is expected to be higher than that of their parent stars,\footnote{This is true in our solar system. Using the standard bulk abundance notation ($X$: hydrogen, $Y$: helium, $Z$: metals), $(Z/X)=0.04-0.12$ for Jupiter \citep{ramirez11} while $(Z/X)_\odot=0.018$ for the Sun \citep{asplund09:review}.} the end result of their formation is a decrease of the metallicity of the host star. Indeed, \cite{ramirez11} found that all elements are deficient by about the same amount in 16\,Cygni\,B relative to 16\,Cygni\,A. More recently, \cite{tucci-maia14} have shown that there is in fact a very small deficiency of refractory elements on top of that constant metallicity offset, which could be the signature of the gas giant planet's rocky core.

The prospect of detecting and characterizing signatures of planet formation using stellar chemical abundances is tantalizing. Few other observations allow us to probe into these processes, and they tend to be costly, in addition to being limited to a handful of very bright and nearby young stars.

In this paper, we investigate the chemical composition of a binary system known to host planets, XO-2, in order to provide further constrains to our hypothesis that planet formation imprints signatures in the chemical composition of the host stars. As in our 16\,Cygni experiment, we begin with the assumption that the two stars in this binary system formed from the same gas cloud, which had a homogeneous chemical composition. Any differences seen in the photospheres of the stars today can then be attributed to processes that occurred either during the planet formation stage or at any other time during the main-sequence lives of the stars. As far as we know, the only previous studies dealing with detailed multi-element chemical abundance analyses of this system are those by \cite{teske13,teske15}.

\section{THE XO-2 STARS AND THEIR PLANETARY SYSTEMS} \label{s:xo-2}

XO-2 (TYC 3413-5-1) is a twin-star binary system where planets have been detected around both components. The stars are both late G-type or early K-type dwarfs of super-solar metallicity. They are separated by  $\simeq30$\,arcsec, which corresponds to a projected distance of about 4\,500\,AU. The difference in effective temperature between them is around 60\,K, making this system comparable to 16\,Cygni in terms of the striking similarity between the two stars. This gives us confidence that a reliable high-precision relative chemical abundance analysis can be applied to them.

\cite{burke07} first announced the discovery of a transiting planet around XO-2N (the ``North'' component of the binary). Further follow-up of the system has revealed the presence of two planets around XO-2S (the ``South'' component; \citealt{desidera14}) as well as evidence of long-term variability in the radial velocities (RVs) of XO-2N \citep[][hereafter D15]{damasso15}. The latter could be interpreted as an additional planet around XO-2N, but the detailed analysis of stellar activity indices by D15 suggests that this variation is likely due to the magnetic cycle of XO-2N instead. According to these authors, who have performed the most recent comprehensive analysis of the XO-2 system and its planets, the mass of the transiting planet orbiting XO-2N is about $0.6\,M_\mathrm{J}$, where $M_\mathrm{J}$ is the mass of Jupiter, while the masses of the planets orbiting XO-2S, which do not transit, are $>0.26\,M_\mathrm{J}$ and $>1.4\,M_\mathrm{J}$. If the long-term RV variations detected in XO-2N are due to another planet, that object would have a mass $>2.4\,M_\mathrm{J}$.

Clearly, the planet population around XO-2 is more complex than the one around 16\,Cygni. In the latter, only one planet of mass $>1.5\,M_\mathrm{J}$ is known to orbit one the stars, namely the secondary 16\,Cygni\,B. The primary has not yet shown any evidence of substellar mass companions \citep{cochran97}. This made the interpretation of the observed chemical abundance anomalies in 16\,Cygni by \cite{ramirez11} and \cite{tucci-maia14} relatively straightforward. XO-2S has two known planets with a total mass $>1.66\,M_\mathrm{J}$. This value could be higher or lower than the total mass of planets around XO-2N depending on the real nature of the long-term RV variability of the latter. And of course we should always be cautious about the fact that smaller planets may have formed around both of these systems, in one or in both of their components, but remain undetected.

\section{SPECTROSCOPIC DATA}

Spectroscopic observations of the XO-2 system were carried out with the HIRES spectrograph on Keck's~10\,m Telescope on 24--25 January 2015. We used the {\tt kv408} filter, which allows a wavelength coverage from about 4080 to 8300\,\AA. The slit width was set to 0.57 arcsec, corresponding to a spectral resolution $R=67\,000$. Four 20 minute exposures for each star were taken, leading to a total $S/N\simeq350$ per pixel in the 5000--7500\,\AA\ region. We also took a solar spectrum of very high $S/N$ ($>700$ per pixel) at $R=84\,000$. Spectra were reduced using the {\tt Keck-MAKEE} standard pipeline which performs bias subtraction, flat-fielding, scattered-light subtraction, 1D spectral extraction and wavelength calibration.

Continuum-normalization and merging of spectral orders (hereafter referred to as ``apertures'') were performed using IRAF's tasks {\tt continuum} and {\tt scopy}. 
A different polynomial order and pixel sampling was employed for each aperture in order to visually confirm the quality of the continuum-tracing in each star. To ensure star-to-star consistency, we adopted the same polynomial orders and pixel sampling for every aperture of the three (previously co-added) spectra available.

\section{ANALYSIS}

\subsection{Input data and analysis tools}

Elemental abundances, including those for iron, which are also used to constrain the relative atmospheric parameters, were determined using a curve-of-growth approach. Specifically, line strength measurements (equivalent widths) were translated into abundances in the stellar photospheres using standard 1D-LTE model atmospheres. The latter were interpolated linearly within the Kurucz {\tt odfnew} grid \cite[e.g.,][]{castelli03}. The {\tt abfind} driver in the 2014 version of the code MOOG \citep{sneden73,sobeck11} was employed for the spectral line calculations.

Equivalent widths ($EW$s) of spectral lines were measured by fitting Gaussian functions to the observed line profiles using IRAF's {\tt splot} task. The majority of lines in our linelist, described in more detail later, are unblended. Those which are somewhat blended typically correspond to features due to elements for which only a handful of spectral lines are available.

Our chemical abundance results can be fully reproduced using the {\tt Qoyllur-quipu} ($q^2$) Python package.\footnote{https://github.com/astroChasqui/q2} Tables provided in this paper contain our $EW$ measurements, and the Keck spectra we employed to make those measurements are available upon request.


\subsection{Atmospheric parameters from an iron-line-only analysis and using a solar spectrum as reference} \label{s:solar}

Atmospheric parameters $\teff$ (effective temperature), $\logg$ (surface gravity), $\feh$ (iron abundance), and $\vt$ (microturbulent velocity) can be measured using only the observed spectra. By measuring line-by-line differential iron abundances from \fei\ and \feii\ lines and minimizing the correlations between relative abundance and excitation potential and reduced equivalent width of the lines ($REW=\log EW / \lambda$), which can be achieved by iteratively fine-tuning the atmospheric parameters themselves, it is possible to find a set of values that satisfy the so-called excitation and ionization balance conditions. This procedure is described in detail in \cite{ramirez14:harps}, where an explanation of the formal error determination can also be found. In particular, we note that the error in the iron abundance inferred is not just the 1\,$\sigma$ line-to-line abundance scatter obtained with the final solution for the parameters, but the result of a rigorous error propagation analysis which is provided for example in \citet[][their Section~3.2]{epstein10} and \citet[][their Appendix~B]{bensby14}.

\begin{figure}
\centering
\includegraphics[width=8cm]{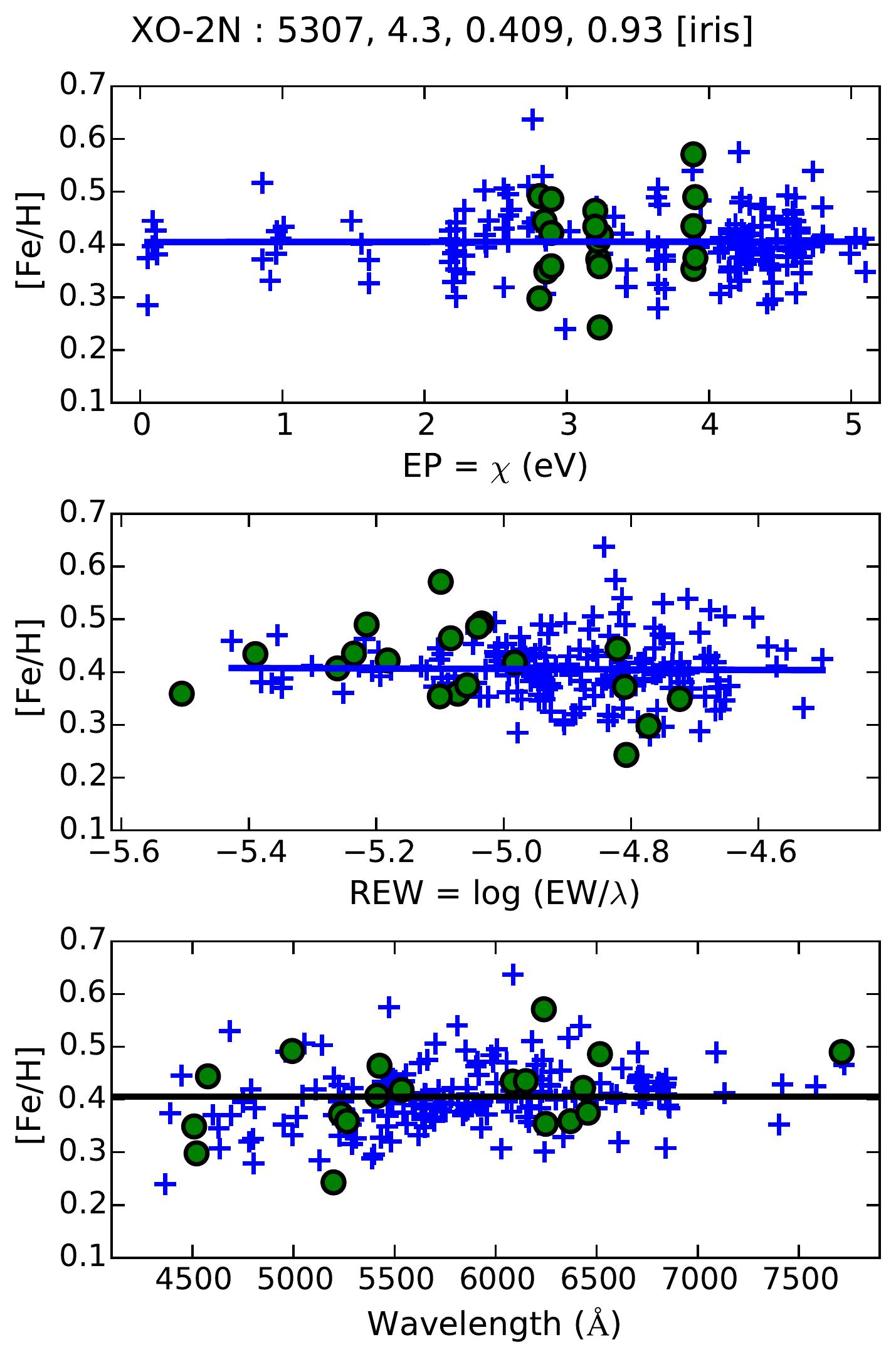}
\caption{Iron abundances measured in the spectrum of XO-2N relative to the solar abundances as a function of the spectral lines' excitation potential (top panel), reduced equivalent width (middle panel), and wavelength (bottom panel). Blue crosses (green circles) correspond to \fei\ (\feii) lines. In the top and middle panels, solid lines are linear fits to the \fei\ data. In the lower panel, the solid line is at the average $\feh$ value.}
\label{f:solarRefN}
\end{figure}

\begin{figure}
\centering
\includegraphics[width=8cm]{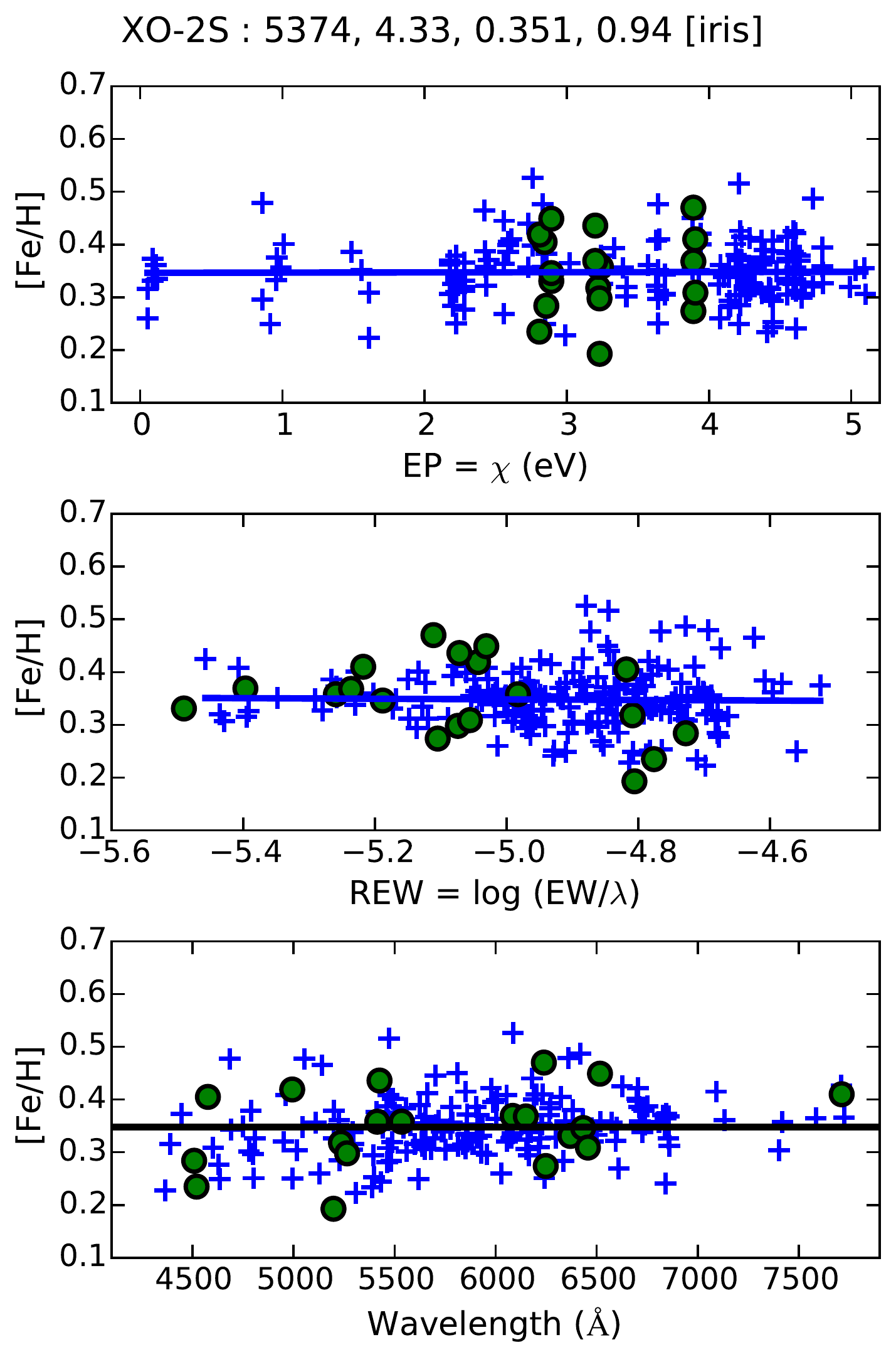}
\caption{As in Figure~\ref{f:solarRefN} for XO-2S.}
\label{f:solarRefS}
\end{figure}

\begin{table*}
\centering
\caption{Atmospheric Parameters}
\begin{tabular}{lcccc}\hline\hline
Star & $\teff$ & $\logg$ & $\feh$ & $\vt$ \\
& (K) & [cgs] & (dex) & (\kms) \\ \hline
\multicolumn{5}{c}{\bf \cite{damasso15}} \\
XO-2N & $5290\pm18$ & $4.43\pm0.10$ & $0.37\pm0.07$ & $0.86\pm0.06$ \\
XO-2S & $5325\pm37$ & $4.42\pm0.10$ & $0.32\pm0.08$ & $0.93\pm0.03$ \\
$\Delta$(N$-$S) & $-35\pm8$ & $+0.01\pm0.02$ & $+0.054\pm0.013$ & $-0.07\pm0.07$ \\ \hline
\multicolumn{5}{c}{\bf Iron-line-only with solar reference (Section~\ref{s:solar})} \\
XO-2N & $5307\pm19$ & $4.30\pm0.05$ & $0.41\pm0.02$ & $0.93\pm0.05$ \\
XO-2S & $5374\pm16$ & $4.33\pm0.04$ & $0.35\pm0.02$ & $0.94\pm0.04$ \\
$\Delta$(N$-$S) & $-67\pm25$ & $-0.03\pm0.06$ & $+0.058\pm0.023$ & $-0.01\pm0.06$ \\ \hline
\multicolumn{5}{c}{\bf Strict N$-$S comparison, IRFM $\teff$ and ``high'' $\logg$ (Section~\ref{s:adopted})} \\
XO-2N & $5440$ & $4.43$ & $0.40$ & $0.94$ \\
XO-2S & $5500\pm5$ & $4.45\pm0.02$ & $0.347\pm0.004$ & $0.94\pm0.02$ \\
$\Delta$(N$-$S) & $-60\pm5$ & $-0.02\pm0.02$ & $+0.053\pm0.004$ & $0.00\pm0.02$ \\ \hline
\end{tabular}
\label{t:pars}
\end{table*}

We employed this method to determine stellar parameters for the XO-2 stars using our solar spectrum as reference, mainly in order to minimize the impact of the uncertainties in the transition probabilities adopted. As initial parameters we employed those obtained in the ``Differential analysis'' of D15 (their Table\,3, for convenience copied here in the first block of our Table~\ref{t:pars}). For the Sun, we adopted $\teff=5777$\,K, $\logg=4.44$, $\feh=0$, and $\vt=1.1$\,\kms, as in D15. Our final results, with the converged parameters, are shown in Figures~\ref{f:solarRefN} and \ref{f:solarRefS}, and the derived parameters with their errors are listed in the second block of Table~\ref{t:pars}. The method used to derive these parameters ensures that the slopes of $\feh$ versus excitation potential and $REW$ are zero within error, but it does not require a zero slope for the $\feh$ versus wavelength relation. Indeed, the latter slopes for Figures~\ref{f:solarRefN} and \ref{f:solarRefS} are not negligible; they are about $+0.02\pm0.01$\,dex per 1\,000\,\AA. This could be related to systematic effects on the continuum fluxes predicted by model atmospheres or due to difficulties in measuring equivalent widths in the spectra of very metal-rich stars relative to the Sun. Indeed, this wavelength trend is closer to zero when the XO-2 stars are directly compared to each other, as will be shown in Section~\ref{s:adopted}.

Figures~\ref{f:solarRefN} and \ref{f:solarRefS} illustrate some of the most important features of our analysis. Note the large number of \fei\ and in particular \feii\ features employed. Note also that the excitation potential coverage of our \fei\ linelist is such that an important number of low excitation potential (EP) \fei\ features are available. These lines are very important for constraining the effective temperature of the star within the iron-line-only analysis because the EP trend is highly sensitive to $\teff$. Other works tend to have few of those lines because they are difficult to measure (many low EP \fei\ lines require very careful de-blending). Instead of rejecting those lines, we make an effort to measure as many of them as precisely as possible.

The parameters we derived as described in this section are in reasonably good agreement with those by D15. Our $\teff$ values are slightly warmer, more so for XO-2S than XO-2N, which implies a larger N--S $\teff$ difference in our case ($-67$ instead of $-35$\,K). Also, while D15 suggest a $\logg\simeq4.42$ for both stars, we find lower values ($\logg\simeq4.32$). This discrepancy is in principle consistent within the error bars for each individual star, but since we detect about the same offset for both stars, it is unlikely that this difference is just a product of observational noise. Although we do not require extremely accurate $\logg$ values in order to measure very precise relative abundances, they are important for determining the evolutionary state and age of the stars. Thus, in Section~\ref{s:logg} we continue our discussion of the $\logg$ values.

Figures~\ref{f:solarRefN} and \ref{f:solarRefS} show that the iron abundance of XO-2N is slightly higher than that of XO-2S. The iron abundance difference (N$-$S) we find with this procedure is $+0.06\pm0.02$, in good agreement with D15: $+0.05\pm0.01$ ($+0.05\pm0.11$ when using a solar spectrum as reference).

\subsection{Photometric effective temperatures}


Effective temperatures of stars can also be measured using photometric methods. Since they rely on broad-band features rather than spectral lines, they are expected to be less sensitive to model uncertainties than iron-line-only analyses. One of the most powerful photometric effective temperature determination techniques is the so-called InfraRed Flux Method \cite[IRFM; e.g.,][]{blackwell77,ramirez05a,casagrande06,casagrande10}.

Using the photometric data provided in Table\,1 of the D15 paper, and adopting their $E(B-V)=0.019$ value, we computed IRFM effective temperatures for the XO-2 stars assuming $\feh=+0.4$ and $\logg=4.5$ for both stars. The IRFM is very weakly dependent on $\feh$ and $\logg$ so our use of these approximate values is acceptable. The IRFM implementation we adopted is that by \cite{casagrande10}. In this scheme, three temperature values are determined from the $J$, $H$, and $K_S$ bands, respectively. The standard deviation of these three values gives us a first estimate of the $\teff$ error. To account for conservative uncertainties of about 0.1\,dex in $\feh$ and 0.005\,mag in $E(B-V)$, we added in quadrature errors of 10\,K and 30\,K, respectively, to the IRFM $\teff$ error. The $\teff$ values obtained using the IRFM directly, and their errors, are listed in the second column of Table~\ref{t:irfm} (``Direct IRFM'').

The stars' colors can also be employed to measure effective temperatures by means of $\teff$-color calibrations such as those by \cite{casagrande10}. Since these calibrations are based on IRFM $\teff$ values determined in the same manner we calculated the IRFM temperatures of the XO-2 stars above, using $\teff$-color calibrations allows us to reduce the impact of errors in the photometric data on the final, averaged $\teff$ values, while keeping them consistent with the same IRFM effective temperature scale. The latter has been shown to be in good agreement with $\teff$ scales based on direct measurements of angular diameters and bolometric fluxes of a selected number of nearby dwarf and subgiant stars \citep{casagrande10,boyajian13}, so it is desirable to remain consistent with that $\teff$ scale. We calculated $(B-V)$, $(V-J)$, $(V-H)$, and $(V-K_S)$ colors for the XO-2 stars using the photometric data from Table\,1 in D15, and adopting the extinction ratios, $k=E(\mathrm{color})/E(B-V)$, from \cite{ramirez05a}. The average $\teff$ values obtained using IRFM $\teff$-color calibrations are listed in the third column of Table~\ref{t:irfm}. In this table we also provide the weighted averages of direct-IRFM and $\teff$-color values, which we adopt as our final photometric effective temperatures (fourth column of Table~\ref{t:irfm}).

\begin{table}
\centering
\caption{Photometric Effective Temperatures}
\begin{tabular}{lccc}\hline\hline
Star & Direct IRFM & Color calibrations & Weighted average \\
& (K) & (K) & (K) \\ \hline
XO-2N & $5472\pm48$ & $5414\pm56$ & $5447\pm36$ \\
XO-2S & $5530\pm42$ & $5474\pm50$ & $5507\pm32$ \\
$\Delta$(N$-$S) & $-58\pm64$ &  $-60\pm75$ & $-60\pm48$ \\
$\Delta$(N$-$S) & --- &  $-57\pm7$\footnote{Using only the differential photometry by \cite{damasso15}} & --- \\ \hline
\end{tabular}
\label{t:irfm}
\end{table}

The photometric temperatures for the XO-2 stars are significantly warmer (about +140\,K) than those obtained in the iron-line-only analysis. The difference is non-negligible, not within the 1-$\sigma$ error bars, and it has the same sign for both stars, suggesting that it is due to systematic errors. Although the photometric $\teff$ values that we have determined are warmer than those we measured with our iron-line-only analysis as well as those from the D15 work, we should note that they are not too far off from the values given in \cite{torres12}, \cite{mortier13}, and \cite{teske15}, for example. The former used spectrum matching to a grid of synthetic spectra while the latter two employed iron-line-only techniques similar to ours.

According to the photometry, the $\teff$ difference between the XO-2 stars (N--S) is $-60\pm48$\,K. Although in principle consistent with the difference reported by D15 ($-35$\,K) within the error bars, the mean value we obtain from our iron-line only analysis is closer ($-67$\,K). Moreover, using the very precise differential photometry provided by D15 (given in the second block of their Table~1) we can also infer a photometric $\teff$ difference which is nearly insensitive to the exact value of the stars' colors. The differential $(B-V)$, $(V-R)$, and $(V-I)$ colors have such small errors (0.002--0.004\,mag) that they translate into photometric $\teff$ difference errors of less than about 10\,K. Averaging the three differential $\teff$ values obtained in this manner we determined a much more precise photometric N--S $\teff$ difference of $-57\pm7$\,K.

\subsection{Surface gravity and evolutionary state} \label{s:logg}

\begin{figure}
\includegraphics[width=8.7cm]{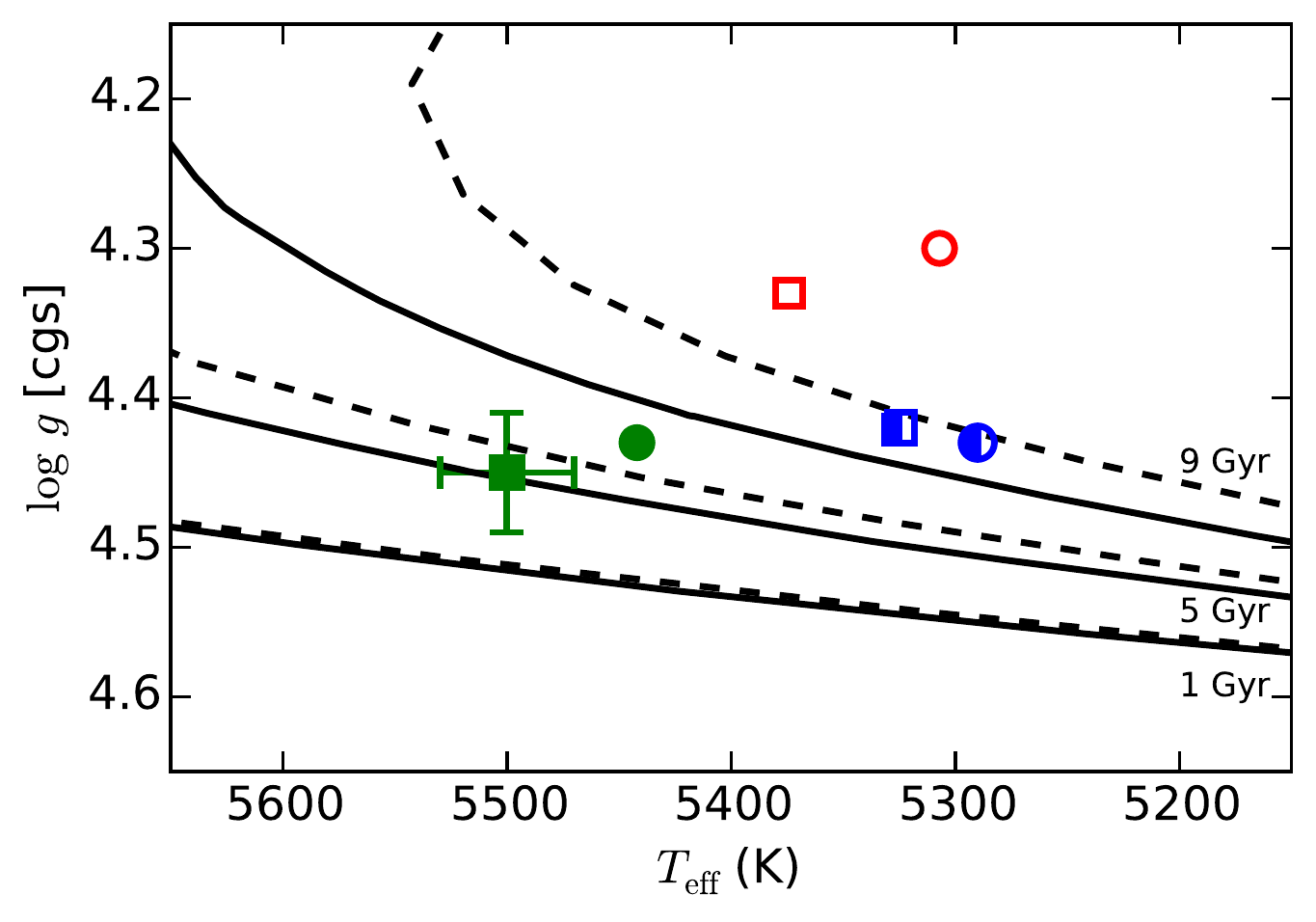}
\caption{Isochrones of 1, 5, and 9\,Gyr as predicted by Yonsei-Yale (dashed lines) and Padova (solid lines) stellar evolution models in the $\teff$-$\logg$ plane. Squares and circles represent the location of the XO-2S and XO-2N stars, respectively, for three different choices of stellar parameter determination: iron-line-only (red open symbols), photometric $\teff$ and  $\logg=4.45$ fixed for XO-2S (green filled symbols), and the values by \cite{damasso15} (blue half-filled symbols). Representative error bars for the  absolute values of $\teff$ and $\logg$ are shown for the data point at $\teff=5500$\,K, $\logg=4.45$.}
\label{f:isochrones}
\end{figure}

As pointed out in Section~\ref{s:solar}, the $\logg$ value that we determined from our iron-line-only analysis is about 0.1\,dex lower than that obtained by D15. To place this important atmospheric parameter in context, in Figure~\ref{f:isochrones} we show theoretical isochrones of 1, 5, and 9\,Gyr (and $\feh=+0.4$) on the $\teff$-$\logg$ plane along with the location of the XO-2 stars depending on the method of atmospheric parameter determination used. Our ``low'' $\logg$ values would make both stars fall outside of the allowed region for Galactic disk stars, i.e., older than about 8\,Gyr (red open symbols). Similarly, the parameters by D15 (blue half-filled symbols) would make the system a very old one ($\gtrsim9$\,Gyr).

If the effective temperatures of the XO-2 stars are closer to 5500\,K, as implied by the photometry, the range of allowed $\logg$ values expands to $\simeq4.3-4.5$. Nevertheless, when we fix the $\teff$ of the XO-2 stars to our photometrically-determined values, therefore basically just forcing the \fei/\feii\ ionization equilibrium, we find that we require $\logg\simeq4.6$ for both stars, which is clearly inconsistent with the predictions of stellar evolution theory. The low levels of chromospheric activity detected in these stars and their long rotation periods suggest a moderately old age. However, the stars are still members of the Galactic thin disk, as argued by D15. Therefore, they must be younger than about 8\,Gyr. Indeed, using the rotation period of XO-2N accurately measured by D15, which is about 42 days, and its $(B-V)$ color, the age of the system according to the gyrochronology formula of \cite{barnes07} is $\simeq6$\,Gyr. For this age (and $\teff\simeq5500$\,K), the preferred $\logg$ values for the XO-2 stars are $\simeq4.45$. By fixing the $\teff$ and $\logg$ values of XO-2S as suggested above, we derived parameters for XO-2N as described in Section~\ref{s:adopted}. These parameters are shown in Figure~\ref{f:isochrones} as green solid symbols.

The parameters from D15 result in an excellent agreement with the isochrones, i.e., both stars fall almost perfectly on the 9\,Gyr Yonsei-Yale isochrone, as expected for a coeval system of two stars. On the other hand, our preferred mean parameters, the green solid symbols, do not fall exactly on the same isochrone. However, considering the uncertainties in absolute stellar parameters (about 30\,K in $\teff$ from the IRFM and about 0.04\,dex in $\logg$ from the iron-line-only analysis relative to the Sun), our preferred values are reasonably consistent with the expectation based on predictions of standard stellar evolution theory.

\subsection{Parameters from a strict N$-$S comparison} \label{s:adopted}

Spectroscopic equilibrium, or excitation/ionization balance of iron lines, is far from being satisfied when employing photometric $\teff$ values and a ``high'' $\logg\simeq4.45$ for the XO-2 stars. The trends with excitation potential and reduced equivalent width are highly significant, as is the difference between iron abundances inferred from \fei\ and \feii\ lines. These trends likely reflect the limitations of traditional 1D-LTE analyses. Future studies with tailored 3D models and accounting for non-LTE effects are of great interest, but beyond the scope of this paper. From a more practical point of view, one can try to avoid these issues by not comparing the XO-2 stars to the Sun anymore and performing a strict differential analysis of XO-2N relative to XO-2S instead, after adopting a set of parameters for XO-2S.

The stellar parameters that we adopt for XO-2S are not entirely arbitrary. Photometric effective temperatures and surface gravities consistent with stellar evolution predictions are more reliable in an absolute sense than $\teff$, $\logg$ combinations that result from purely spectroscopic analyses of iron lines. Thus, hereafter we fix $\teff=5500$\,K, $\logg=4.45$, $\feh=+0.35$, and $\vt=0.94$ for XO-2S, and measure only relative parameters for XO-2N. The solar spectrum is no longer used to calculate stellar parameters. The technique employed to derive the relative parameters is the same one used before with the only exception that XO-2S is the reference star instead of the Sun.

\begin{figure}
\centering
\includegraphics[width=8cm]{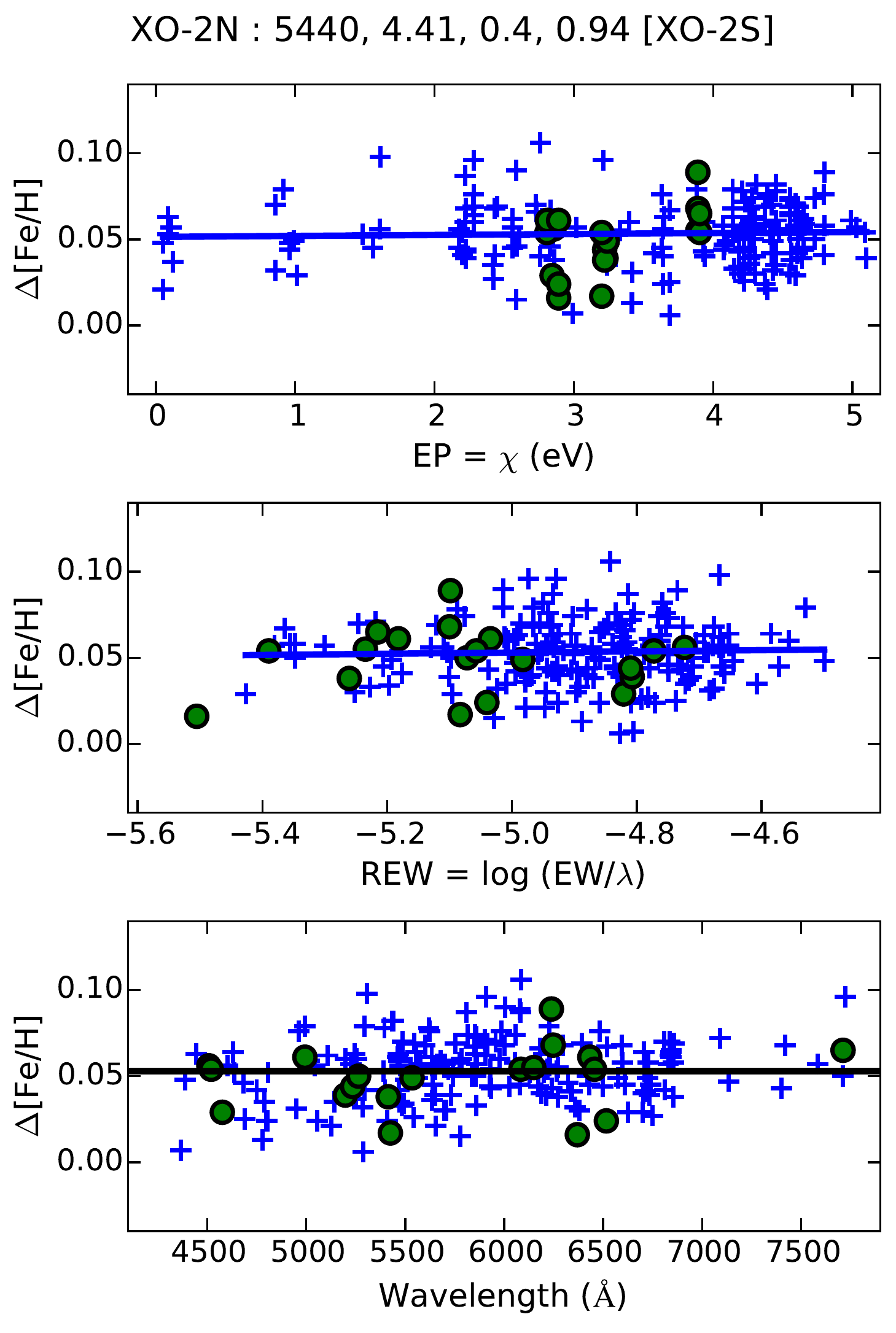}
\caption{As in Figure~\ref{f:solarRefN} for the line-by-line analysis of XO-2N relative to XO-2S. Note that the scale of the vertical axis here is about four times smaller than that in Figures~\ref{f:solarRefN} and \ref{f:solarRefS}, and that the line-to-line scatter here is about six times smaller because the two stars being compared are very similar to each other, minimizing systematic errors.}
\label{f:NS}
\end{figure}

The final result of our strict line-by-line differential stellar parameter determination is shown in Figure~\ref{f:NS}. The derived parameters for XO-2N and their formal errors are listed in the third block of Table~\ref{t:pars}. Note that the latter are considerably smaller than those obtained before. This is due to the fact that we are now comparing two stars which are twins of each other, and do not use a solar spectrum as an intermediate step.\footnote{Since the $EW$s for the iron lines were measured ``manually'' by one of us, instead of using an automated code, one could argue that our results are subjective. To address this potential problem, another set of iron line $EW$s were measured by a different person and the differential parameters were determined using this independent set of $EW$s. In that case we found $\Delta\teff=-65\pm8$\,K, $\Delta\logg=-0.03\pm0.03$, $\Delta\feh=+0.063\pm0.010$, and $\Delta\vt=-0.03\pm0.03$, in very good agreement with the first case, albeit with somewhat larger error bars. This shows that with very high-quality spectra the subjectivity of manual $EW$ measurements is not so important.} In Section~\ref{s:intro}, we advocated using the Sun as a comparison. This approach is of great value when analyzing solar twins. Comparing abundances of XO-2N and XO-2S with each other rather than the Sun significantly reduces systematic errors. Indeed, as noted in Section~\ref{s:solar}, a non-negligible slope for the $\feh$ versus wavelength data is observed when comparing these stars to the Sun. The corresponding slope for Figure~\ref{f:NS} ($\Delta\feh$ versus wavelength) is about four times smaller: $+0.005\pm0.002$\,dex per 1\,000\,\AA, yet still not zero within the very small errors in this twin-star comparison.

Our strict N--S comparison, in which we adopted a photometric (IRFM) effective temperature and a ``high'' surface gravity for the reference star XO-2S, reveals a very precise effective temperature difference (N--S) of $-60\pm5$\,K, in excellent agreement with the very precise value obtained from the differential photometry ($-57\pm7$\,K). Both are significantly higher than the difference measured by D15 ($-35\pm8$\,K).

Interestingly, despite the differences in stellar parameters adopted and the $\teff$ difference, the offset in $\feh$ between XO-2N and XO-2S is almost exactly the same: $+0.054\pm0.013$ according to D15 versus our $+0.053\pm0.004$. In fact, any of the other combinations of stellar parameters that we have found so far always result in an iron-enhanced XO-2N relative to XO-2S by at least +0.05\,dex. There is little doubt that the iron abundances of these two stars are different.

According to our strict N--S comparison, $\Delta\logg=-0.02\pm0.02$ (N--S). This means that the cooler N component has a slightly lower $\logg$, which implies an isochrone age about 1\,Gyr older than the S component, in contradiction with the assumption of coeval age for stars in binary systems. Note, however, that the difference is marginally consistent with zero, and in fact a detailed isochrone age calculation (as in \citealt{ramirez14:harps}) results in error bars of about 1\,Gyr for each star. Nevertheless, it is interesting to note that adopting a $\logg=4.45$ for XO-2N results in an \feii\ minus \fei\ iron abundance difference of +0.01\,dex, as opposed to exactly zero. This slightly higher \feii\ abundance could be related to the \feii\ minus \fei\ discrepancies seen in the Hyades cluster by \cite{yong04}, who employed $\logg$ values determined from isochrones and a photometric $\teff$ scale. Thus, it is possible that the ``peculiar'' $\logg$ value of XO-2N is again a product of model deficiencies. However, as we discuss later in this paper, this issue does not affect our relative abundance ratios in any significant way.

\subsection{Elemental abundance determination}

Elemental abundances of 22 elements other than iron were measured for the XO-2 stars. For carbon, we used C\,\textsc{i} and CH lines, while for Sc, Ti, and Cr, spectral lines due to the neutral and singly-ionized species were employed. The linelist and atomic parameters adopted are the same as in \cite{ramirez14:harps}, except for the CH and Rb lines. For these species we adopted the line data compiled by \cite{asplund05:solarabundances} and \cite{grevesse15}, respectively. Hyperfine structure was taken into account for V, Mn, Co, Cu, Rb, Y, and Ba. The linelist and atomic parameters used in this work are listed in Table~\ref{t:linelist} along with the EWs measured in the XO-2 stars and our solar spectrum. The relative abundances obtained from each species, and their errors, are given in Table~\ref{t:abundances}. The oxygen abundances inferred from the O\,\textsc{i} 777\,nm triplet were corrected for differential non-LTE effects using the grid and code by \cite{ramirez07}.

\begin{table}
\centering
\caption{Line List}
\begin{tabular}{lrllrrr}\hline\hline
Wavelength & Species & EP & $\log gf$ & XO-2N & XO-2S & Sun \\
(\AA) & & (eV) & & (m\AA) & (m\AA) & (m\AA) \\
\hline
5052.167 & 6.0 & 7.685 & -1.304 & 32.0 & 34.2 & 32.8 \\
6587.6099 & 6.0 & 8.537 & -1.021 & 10.8 & 12.1 & 13.8 \\
7111.469 & 6.0 & 8.64 & -1.074 & 17.7 & 19.1 & 10.7 \\
7116.96 & 6.0 & 8.65 & -0.91 & 11.9 & 12.7 & 13.2 \\
4218.723 & 106.0 & 0.413 & -1.008 & 92.5 & 93.9 & 79.5 \\
4253.003 & 106.0 & 0.523 & -1.523 & 43.3 & 44.3 & 33.3 \\
4253.209 & 106.0 & 0.523 & -1.486 & 42.2 & 43.8 & 33.8 \\
7771.9438 & 8.0 & 9.146 & 0.352 & 44.2 & 48.3 & 68.7 \\
7775.3901 & 8.0 & 9.146 & 0.002 & 38.9 & 42.0 & 48.0 \\
4751.8218 & 11.0 & 2.104 & -2.078 & 39.1 & 34.9 & 18.4 \\
4982.82 & 11.0 & 2.1 & -1.0 & 103.5 & 101.9 & 81.9 \\
5148.838 & 11.0 & 2.102 & -2.044 & 40.9 & 36.3 & 10.0 \\

\vdots & \vdots & \vdots & \vdots & \vdots & \vdots & \vdots \\
\hline
\end{tabular}
\label{t:linelist}
\end{table}

\begin{table}
\centering
\caption{Relative Abundances}
\begin{tabular}{lrcc}\hline\hline
Species & $\tc$ & $\Delta$[X/H] & error \\
& (K) & (dex) & (dex) \\
\hline
C\,\textsc{i}   &   40 & 0.009 & 0.009 \\
CH   &   40 & 0.015 & 0.006 \\
O\,\textsc{i}   &  180 & 0.013 & 0.008 \\
Na\,\textsc{i}  &  958 & 0.035 & 0.009 \\
Mg\,\textsc{i}  & 1336 & 0.074 & 0.012 \\
Al\,\textsc{i}  & 1653 & 0.087 & 0.018 \\
Si\,\textsc{i}  & 1310 & 0.070 & 0.006 \\
S\,\textsc{i}   &  664 & 0.024 & 0.008 \\
K\,\textsc{i}   & 1006 & 0.008 & 0.027 \\
Ca\,\textsc{i}  & 1517 & 0.071 & 0.009 \\
Sc\,\textsc{i}  & 1659 & 0.102 & 0.024 \\
Sc\,\textsc{ii} & 1659 & 0.076 & 0.010 \\
Ti\,\textsc{i}  & 1582 & 0.085 & 0.009 \\
Ti\,\textsc{ii} & 1582 & 0.082 & 0.010 \\
V\,\textsc{i}   & 1429 & 0.078 & 0.009 \\
Cr\,\textsc{i}  & 1296 & 0.044 & 0.008 \\
Cr\,\textsc{ii} & 1296 & 0.067 & 0.011 \\
Mn\,\textsc{i}  & 1158 & 0.035 & 0.008 \\
Fe\,\textsc{i}  & 1334 & 0.054 & 0.005 \\
Fe\,\textsc{ii} & 1334 & 0.058 & 0.007 \\
Co\,\textsc{i}  & 1352 & 0.058 & 0.006 \\
Ni\,\textsc{i}  & 1353 & 0.061 & 0.006 \\
Cu\,\textsc{i}  & 1037 & 0.055 & 0.011 \\
Zn\,\textsc{i}  &  726 & 0.020 & 0.006 \\
Rb\,\textsc{i}  &  800 & 0.019 & 0.017 \\
Y\,\textsc{ii}  & 1659 & 0.082 & 0.019 \\
Zr\,\textsc{ii} & 1741 & 0.116 & 0.026 \\
Ba\,\textsc{ii} & 1455 & 0.074 & 0.017 \\ \hline
\end{tabular}
\label{t:abundances}
\end{table}

For each species, we calculated average values of $\Delta$[X/H](N--S) and the standard deviation of the line-to-line scatter ($\sigma$), from which a standard error was computed as $\delta_\mathrm{lines}=\sigma/\sqrt{n-1}$, where $n$ is the number of spectral lines employed. In cases where only one spectral line is available we conservatively adopted $\delta_\mathrm{lines}=0.025$\,dex, which corresponds to the largest error obtained for species with more than three lines available. On the other hand, by propagating the formal errors in the relative atmospheric parameters of XO-2N, as given in the last row of Table~\ref{t:pars}, we computed an uncertainty in the average $\Delta$[X/H] values, $\delta_\mathrm{pars}$, which varies from one species to another in the 0.003--0.011\,dex range. We added $\delta_\mathrm{lines}$ and $\delta_\mathrm{pars}$ in quadrature to obtain the final error bars for our $\Delta$[X/H](N--S) relative abundances. 

\begin{figure*}
\centering
\includegraphics[width=18cm]{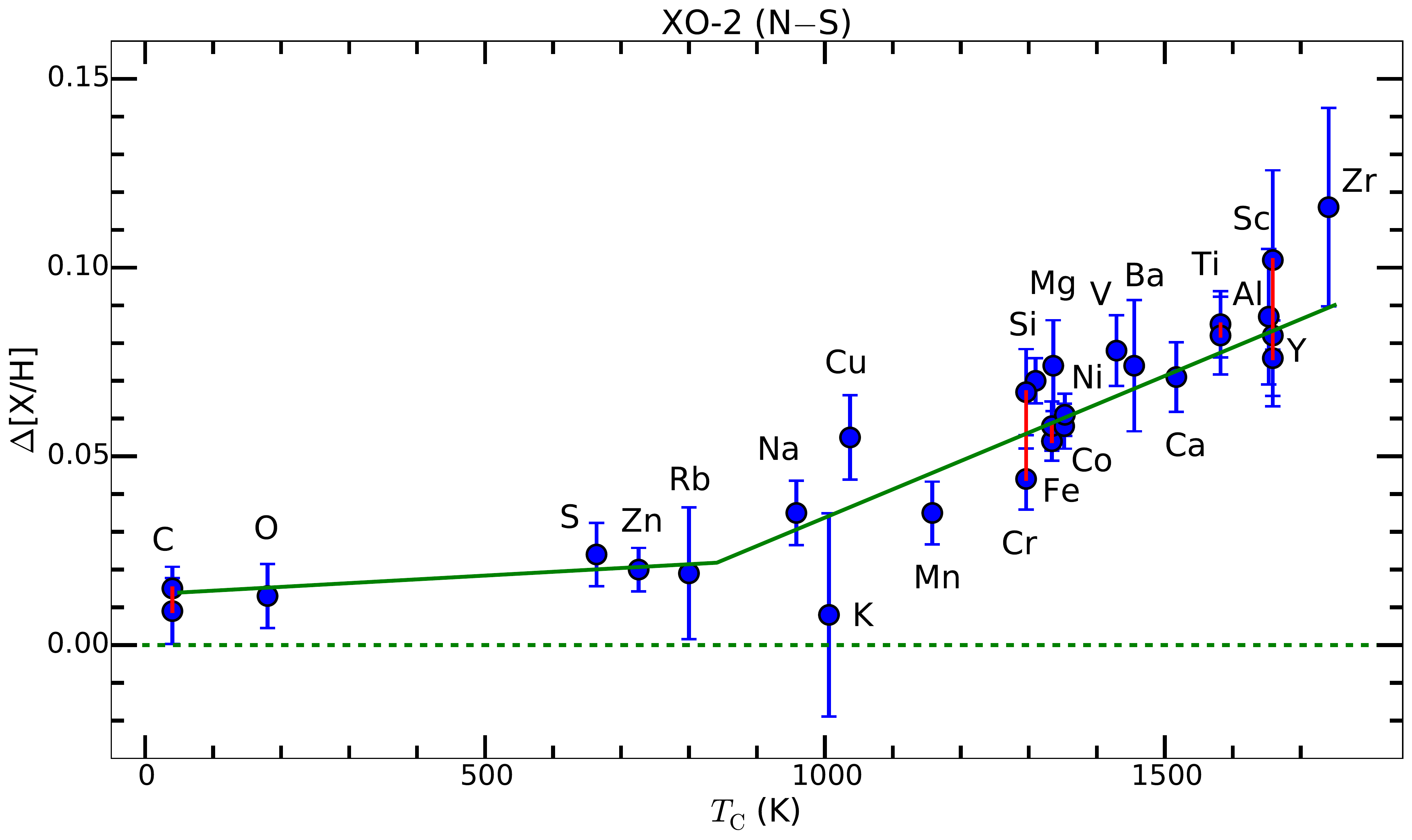}
\caption{Chemical composition difference between XO-2N and XO-2S as a function of the elements' condensation temperature. The dashed line corresponds to identical chemical composition. The solid line is a double linear fit to the data, broken at $\tc=841$\,K. Red vertical lines connect two species of the same chemical element (e.g., CH and C\,\textsc{i}, Ti\,\textsc{i} and Ti\,\textsc{ii}, etc.).}
\label{f:abNS}
\end{figure*}

Figure~\ref{f:abNS} shows the elemental abundance difference between XO-2N and XO-2S as a function of condensation temperature, the latter as given by \citet[][specifically their 50\,\% condensation sequence for a solar-composition mixture]{lodders03}. The correlation between $\Delta$[X/H] and condensation temperature is very strong while the $\simeq+0.05$\,dex offset in the iron abundance previously reported by D15 and confirmed in this work fits the $\tc$ trend perfectly ($\tc=1334$\,K for iron).\footnote{Most, if not all studies that have looked for correlations between abundance anomalies and condensation temperature have employed the solar $\tc$ sequence as reference (i.e., as the $x$-axis in figures like our Figure~\ref{f:abNS}) and for statistics associated with it. However, the $\tc$ values depend on the composition of the gas, and one would expect the differences with the solar case to be particularly important for metal-rich systems like XO-2. Detailed chemical equilibrium calculations to determine the $\tc$ sequences on a star-by-star basis, such as those by \cite{bond10}, are beyond the scope of this paper. We note, however, that the $\tc$ sequences by \citeauthor{bond10} for stars somewhat similar to those in the XO-2 system result in noisier $\tc$ trends. Although the overall correlation (the ``upward'' trend seen in Figure~\ref{f:abNS}) persists, the $\tc$ values computed by \citeauthor{bond10} for C and O can be significantly higher at high metallicity (up to about 1\,000\,K) while those of some refractories like Si and Ti could be higher by about 300\,K. These effects should be investigated in future works in this area.}

A double linear fit to the $\Delta$[X/H] versus $\tc$ data is also shown in Figure~\ref{f:abNS}. We used the IDL {\tt MPFIT} routine \citep{markwardt09} to find a robust value for the $\tc$ of the break. This routine uses the Levenberg-Marquardt technique for least-squares minimization and requires an initial guess. We tested a large range of initial guesses and found the break to be at $\tc=841\pm137$\,K. With a break at $\tc=841$\,K the connection between the two linear fits is smooth. Each linear fit was calculated using the inverse square of the $\Delta$[X/H] errors as weights. The minimum and maximum $\Delta$[X/H] values of the fit are 0.014 and 0.090\,dex, respectively. The element-to-element scatter around the fit is 0.011\,dex, which is very similar to the average value of our $\Delta$[X/H] errors (0.012\,dex). This suggests that the linear fits shown in Figure~\ref{f:abNS} constitute an excellent representation of the abundance differences between the XO-2N and XO-2S stars and that the element-to-element scatter is due to observational errors alone. Furthermore, this implies that the $\Delta$[X/H] difference for the most volatile elements such as C and O is slightly larger than zero. In fact the volatile element difference can be quantified as $+0.015\pm0.011$\,dex. Since the most refractory elements are over-abundant in XO-2N relative to XO-2S by about 0.090\,dex, the amplitude of the $\tc$ trend shown in Figure~\ref{f:abNS} is about 0.075\,dex.

\subsection{Dependence on adopted stellar parameters}

Here we discuss how the result in Figure~\ref{f:abNS} depends on alternative choices of stellar parameters. 
Throughout this exercise, we have used our $EW$ measurements. Thus, it is not our intention to replicate the results of previous studies, whenever applicable, but rather to test whether the very strong $\Delta$[X/H] versus $\tc$ correlation that we have found is only due to our particular choice of stellar parameters or their N--S difference.

\begin{figure}
\centering
\includegraphics[width=8.6cm]{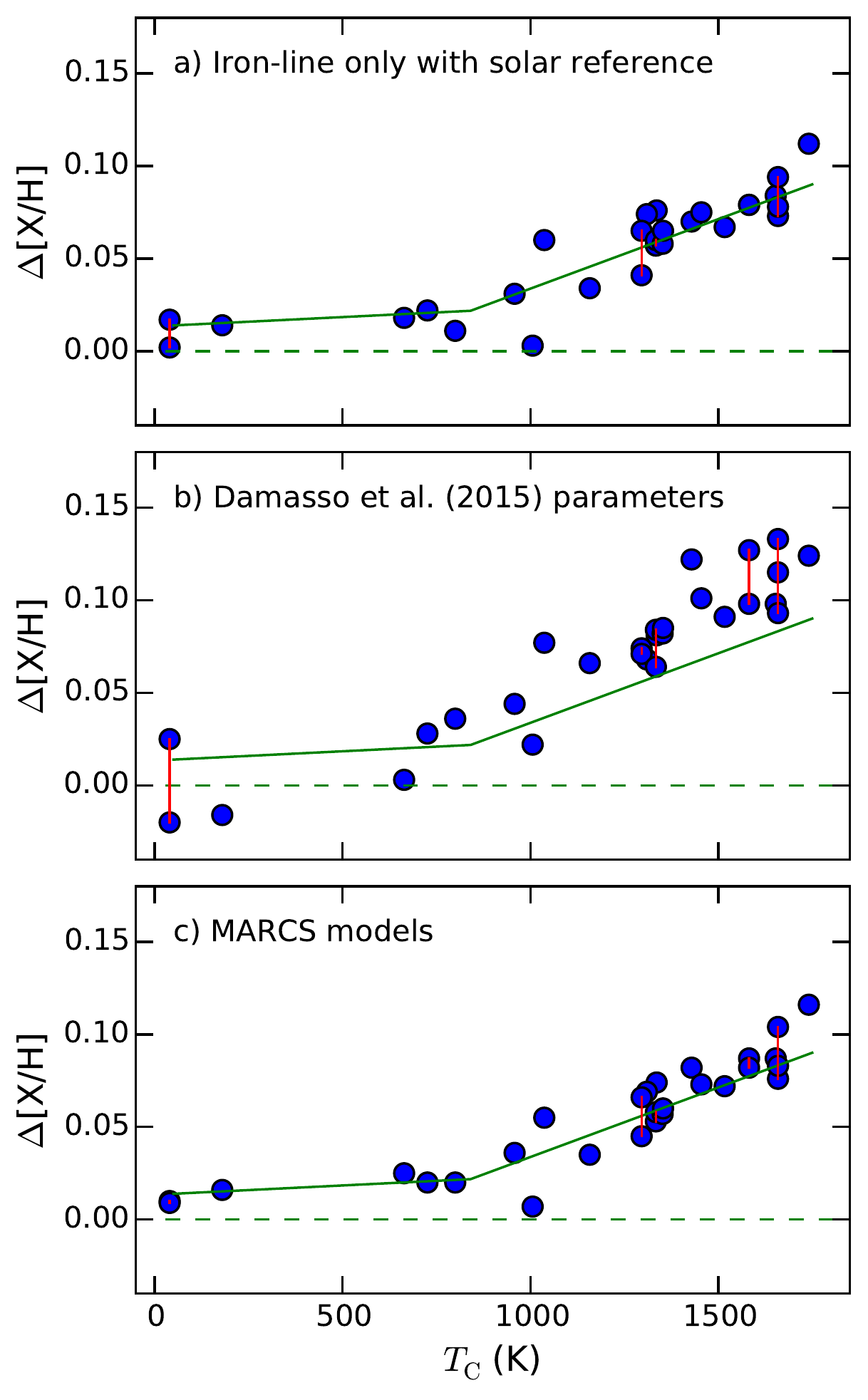}
\caption{As in Figure~\ref{f:abNS}, but for different choices of stellar parameters for XO-2N and XO-2S. The solid lines correspond to the fit to our data in Figure~\ref{f:abNS}.}
\label{f:tc_comp}
\end{figure}

We begin by recalculating the abundances using our iron-line-only parameters from Section~\ref{s:solar}, which employs the Sun as the reference star. These parameters are given in the middle section of Table~\ref{t:pars}. The most important differences with our preferred parameters are the lower $\teff$ and $\logg$ values, but the N--S differences are essentially the same. Panel a) in Figure~\ref{f:tc_comp} shows the XO-2N minus XO-2S abundance difference in this case. The $\tc$ trend is still very clear and there appears to be only a minor offset of about $-0.01$\,dex relative to our preferred abundance set. In this case the volatile element abundances are slightly more similar between XO-2N and XO-2S, but the latter still seems to be slightly volatile poor. Although the $\teff$ values used in this case are about 140\,K cooler than the more reliable IRFM temperatures, and the $\logg$ values are too low (see Section~\ref{s:logg}), the fact that the N--S relative parameters are basically the same leads to a very similar $\Delta$[X/H] versus $\tc$ correlation.

Panel b) in Figure~\ref{f:tc_comp} shows the abundances we derive using the parameters from \cite{damasso15}, which are given in the first section of Table~\ref{t:pars}. Perhaps the most important difference with our parameters is the $\Delta\teff$ value, which is about twice as large in our case (D15 suggest $\Delta\teff=-35$\,K, but we find $\Delta\teff=-60$\,K). Also, the $\teff$ values themselves are offset by about 160\,K, although this might be of secondary importance considering the test discussed in the previous paragraph. 

Using the parameters from D15, the $\Delta$[X/H] versus $\tc$ correlation appears steeper, but it is also more noisy than the one we obtained using our parameters. We note also that in this case certain elements with more than one species available for analysis show abundance differences which are less consistent than with our parameters. For example the carbon abundances obtained from C\,\textsc{i} and CH lines show a difference of 0.04\,dex whereas with our parameters the difference is only 0.006\,dex. Also, the Ti abundance from Ti\,\textsc{i} lines is about 0.03\,dex higher than that from Ti\,\textsc{ii} lines in this case, but we find a difference of 0.003\,dex with our parameters. Based on these results, we argue that our differential parameters are more accurate than those given by D15. In any case, the $\tc$ trend of the elemental abundance difference is not at all removed when using their parameters. In fact, as noted above, the $\tc$ trend seems to be steeper, with the volatile abundances being nearly the same between XO-2N and XO-2S, or even slightly lower, but only if the carbon abundance from the CH lines is ignored.

In panel c) of Figure~\ref{f:tc_comp} we show the relative abundances that we derive when we adopt MARCS model atmospheres instead of those from Kurucz's {\texttt odfnew} grid. The differences are all within 0.01\,dex. Therefore, adopting a different model atmosphere grid has no impact on the $\Delta$[X/H] versus $\tc$ correlation found. As already shown in some of our previous papers \cite[e.g.,][]{ramirez11,melendez12}, the choice of model atmosphere grid is irrelevant when dealing with high-precision ($\simeq0.01$\,dex) relative abundances of stellar twins.

\begin{figure}
\centering
\includegraphics[width=8.7cm]{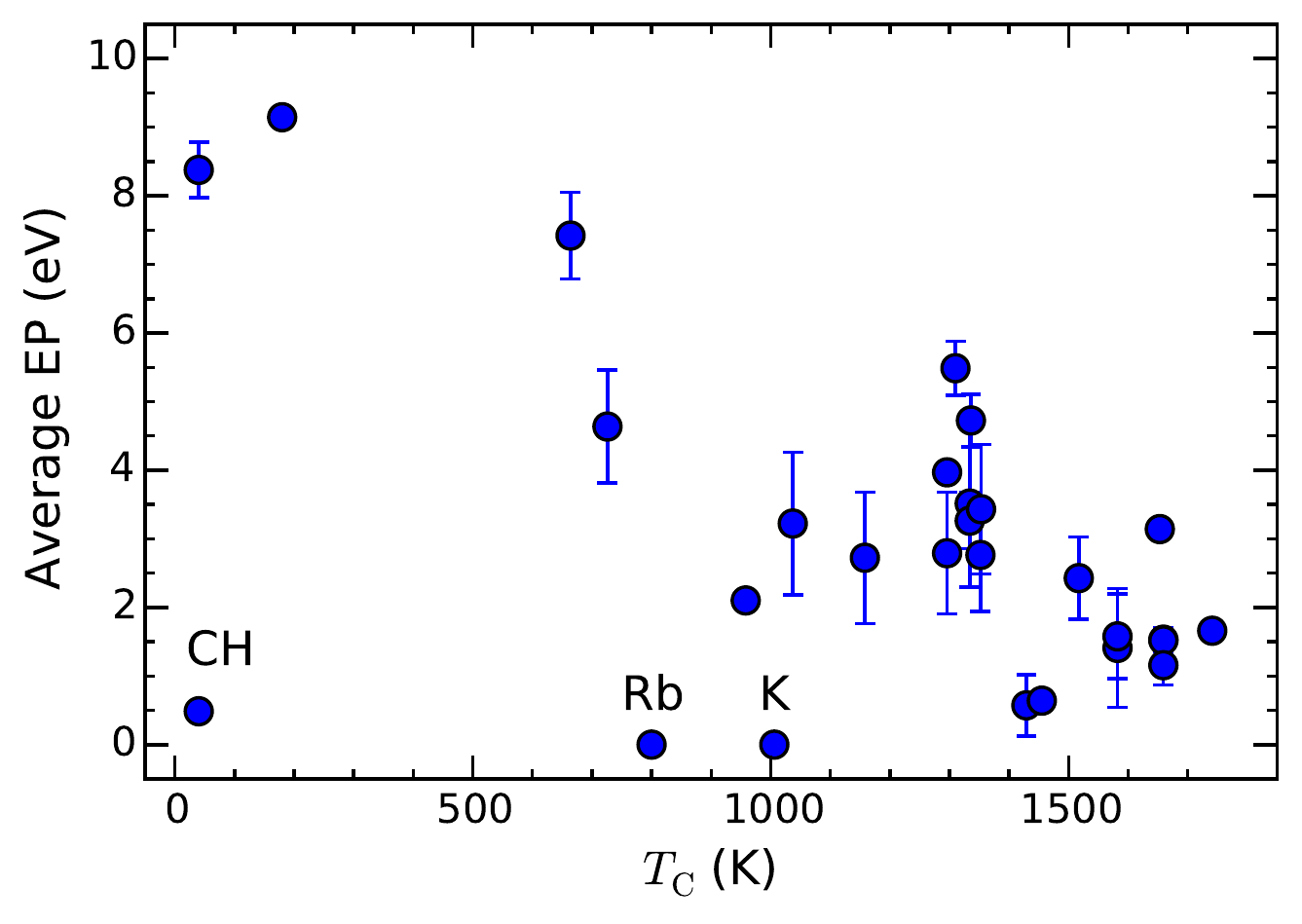}
\caption{Average excitation potential of the spectral lines used in this work as a function of the elements' condensation temperatures. The three species which depart the most from the mean trend (CH, Rb, and K) are labeled.}
\label{f:tc_ep}
\end{figure}

\begin{figure}
\includegraphics[width=8.7cm]{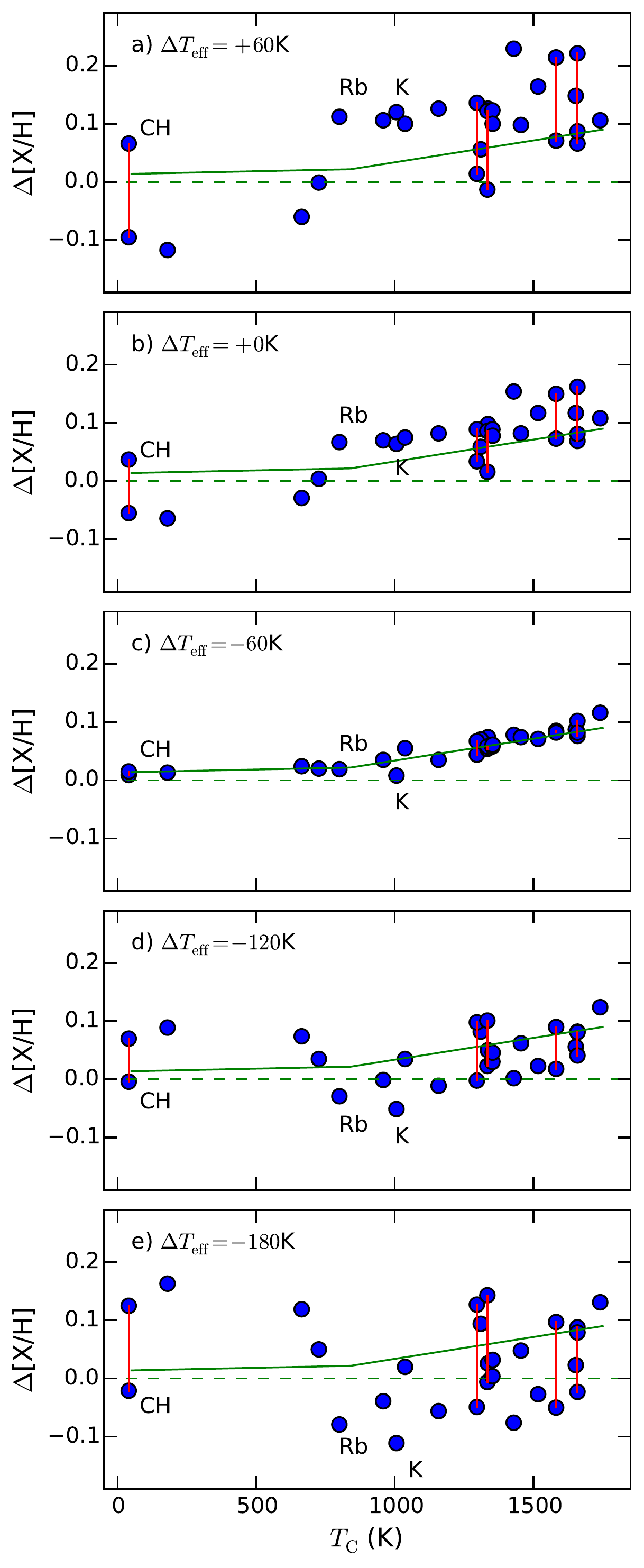}
\caption{As in Figure~\ref{f:abNS}, but for different values of the XO-2N minus XO-2S temperature difference, $\Delta\teff$, keeping all other parameters (from the third block of Table~\ref{t:pars}) constant. Note that panel c) is a replica of Figure~\ref{f:abNS}. The solid lines correspond to the fit to panel c)'s data. The three species that depart the most from the average EP versus $\tc$ correlation (CH, Rb, and K), shown in Figure~\ref{f:tc_ep}, are labeled.}
\label{f:tc_ep_test}
\end{figure}

Based on the discussion above, it appears to be the case that a very precise knowledge of the temperature difference between the XO-2 stars is critical to trace the $\Delta$[X/H] versus $\tc$ correlation. This would be expected if the average excitation potential of the lines employed for each chemical element correlates with the condensation temperature. Figure~\ref{f:tc_ep} shows that this is in fact the case. With few exceptions (CH, Rb, and K), the average EP of the lines used for a given element is correlated with the elements' $\tc$. Thus, one might wonder what $\Delta\teff$ value would completely blur the $\Delta$[X/H] versus $\tc$ correlation and/or flip the sign of the slope.

Figure~\ref{f:tc_ep_test} shows the results obtained for the XO-2N minus XO-2S abundance difference for various arbitrary choices of $\Delta\teff$(N--S). Only the effective temperature difference is changed; all input data as well as the other parameters, including $\teff=5500$\,K for XO-2S, were kept constant. We varied $\Delta\teff$ from +60\,K to $-180$\,K, but note that positive $\Delta\teff$ values are unrealistic given the very precise differential photometry of these objects and the fact that they are co-eval main-sequence stars. XO-2S is definitely brighter and bluer, therefore warmer than XO-2N, which implies $\Delta\teff<0$.

The element-to-element scatter in $\Delta$[X/H] versus $\tc$ increases for both lower and higher values of $\Delta\teff$ relative to our preferred value ($-60$\,K), which corresponds to panel c) in Figure~\ref{f:tc_ep_test}. Moreover, the agreement in abundance difference for elements with two species available is excellent when using $\Delta\teff=-60$\,K, but it deteriorates for other values (see the filled circles connected by red vertical lines in Figure~\ref{f:tc_ep_test}). This is particularly the case of element C, for which high excitation C\,\textsc{i} and low excitation CH lines are available. While the C abundance difference from C\,\textsc{i} lines becomes less negative and eventually more positive for higher values of $\Delta\teff$, the C abundance from CH lines shown the opposite behavior. The C abundance difference from both species is virtually the same for $\Delta\teff=-60$\,K.

Inspection of Figure~\ref{f:tc_ep_test} shows that the $\Delta$[X/H] versus $\tc$ slope remains positive, albeit with progressively larger error, up to $\Delta\teff=+60$\,K and down to $\Delta\teff=-120$\,K, but it changes sign at about $\Delta\teff=-180$\,K, particularly if CH lines are not used. Thus, only if the absolute value of the temperature difference between XO-2N and XO-2S is highly overestimated, by more than 100\,K, would the observed $\tc$ slope change sign. 

Figure~\ref{f:tc_ep} could also be used in the XO-2 context to explore the possibility that stellar activity affects the derived abundances at high precision. D15 have noticed that in the timespan of their observations XO-2N was on average more active than XO-2S. Low excitation lines, thought to form preferentially at high atmospheric layers, would be more affected by activity-related heating than high excitation lines. If this had a significant impact on the line strengths, one would expect it to introduce an abundance trend with EP (and therefore $\tc$).

However, we must note that line formation depth is much more sensitive to line strength than EP \cite[e.g.,][]{gurtovenko15} and that our linelist is extremely heterogeneous in this regard. In addition, while the different activity levels of the XO-2 stars could be explained by different current phases of their activity cycles, one should notice that the observed ranges of activity indices in XO-2N and XO-2S overlap significantly (Figures~15 and 17 in D15). Thus, the probability that the activity levels of the two stars are very similar at any given time of observation is not low. Unfortunately, our spectra do not include the \ion{Ca}{2} H\&K lines.

\subsection{Using parameters from \cite{teske15}}

\citet[][hereafter T15]{teske15} provided an in-depth investigation of stellar parameter dependency for the chemical abundance difference between the stars in the XO-2 system. Four sets of parameters were determined by them (they are listed in their Table~1), from which they calculated abundances of 16 elements using spectra of $R\sim60\,000$ and $S/N\sim170-230$ taken with the High Dispersion Spectrograph at the 8.2\,m Subaru Telescope.

\begin{figure}
\centering
\includegraphics[width=8.6cm]{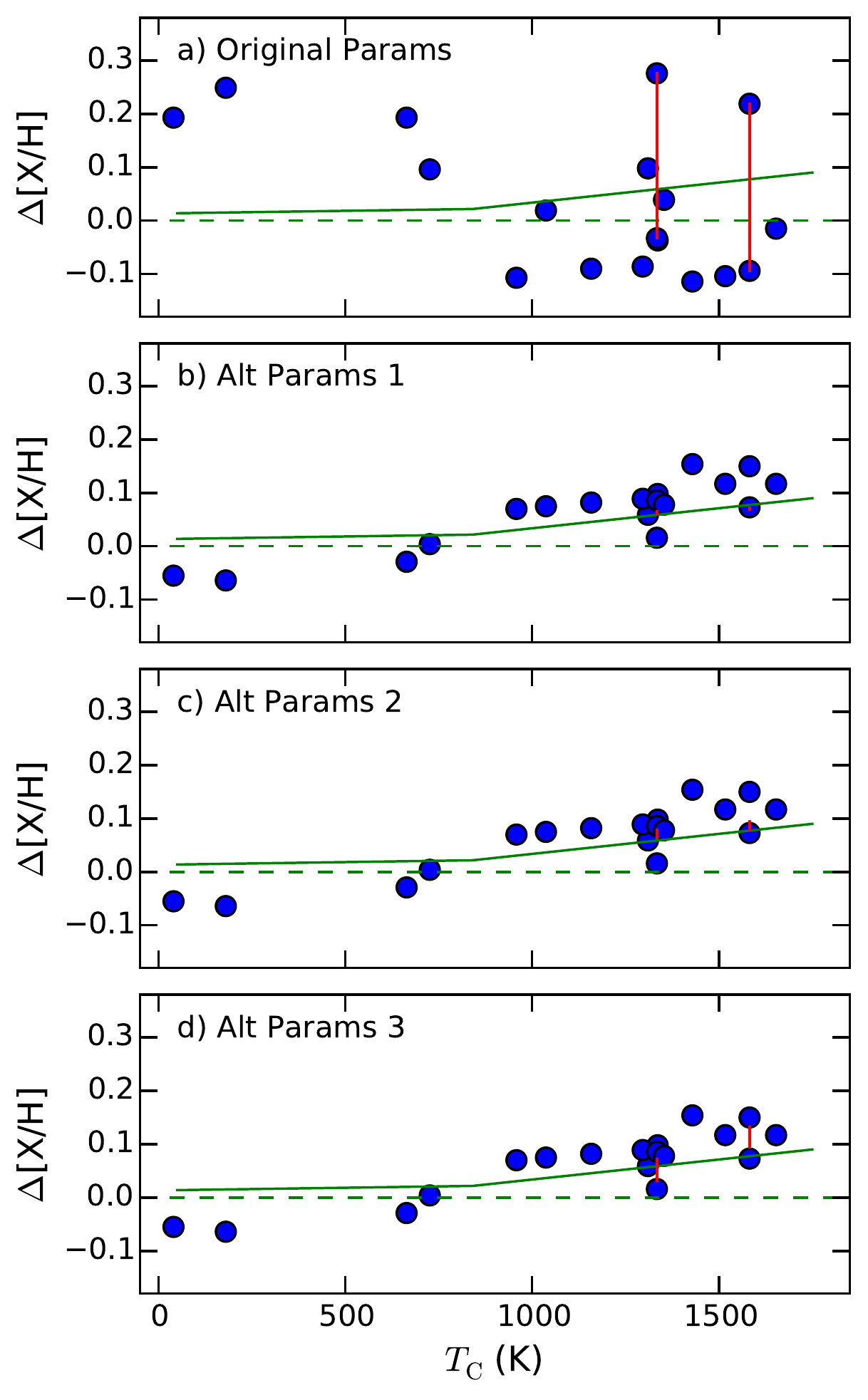}
\caption{As in Figure~\ref{f:abNS}, but using input atmospheric parameters from \cite{teske15} and restricted to the species used in that work. The solid lines correspond to the fit to our data in Figure~\ref{f:abNS}.}
\label{f:tc_teske}
\end{figure}

Figure~\ref{f:tc_teske} shows the $\Delta$[X/H] versus $\tc$ relations we find when we use our $EW$ measurements, but the stellar parameters from T15. We show only the data points that correspond to the species used in T15. Note in particular that their C abundances are based on C\,\textsc{i} lines only. Also, while their O abundances were obtained via spectrum synthesis of the forbidden 630\,nm line, ours were calculated from $EW$ measurements the O\,\textsc{i} triplet at 777\,nm.

When using T15's ``original parameters,'' the $\Delta$[X/H] versus $\tc$ relation shows a negative slope, as shown in panel a) of Figure~\ref{f:tc_teske}. Note that in this case two values of each Fe and Ti abundances are available, and they disagree badly (see the red vertical solid lines). Panel a) of Figure~\ref{f:tc_teske} closely resembles panel e) of Figure~\ref{f:tc_ep_test}. Indeed, the $\Delta\teff$(N--S) according to T15's original parameters is $-204$\,K, much cooler than our precise value of $-60$\,K. In any case, we note that our N--S abundance differences for this set of parameters are in very good agreement with those found by T15. The general appearance of the black asterisks in their Figure~1 closely resembles that of our filled circles in Figure~\ref{f:tc_teske}a). In particular, we note that the $\Delta$[X/H] values for C, O, S, and Zn (all with $\tc<800$\,K), are consistently all positive.

Panels b), c), and d) of Figure~\ref{f:tc_teske} show the $\Delta$[X/H] versus $\tc$ correlation that we obtain using T15's ``alternative parameters'' 1, 2, and 3, respectively. All of them exhibit somewhat steeper slopes compared to our mean trend (shown with green solid lines), but also more element-to-element scatter. The alternative parameters 1 and 2 have $\Delta\teff=-83$\,K and $-72$\,K, values which are not very different from ours. For the alternative parameters set 3, $\Delta\teff=+3$\,K.

While the steeper $\tc$ slope can be explained for the alternative parameter set 3 by comparing it to our test in panel b) of Figure~\ref{f:tc_ep_test}, the steeper slopes and noisier behaviors obtained with the alternative parameters 1 and 2 are more likely due to discrepancies in the other stellar parameters. In any case, it is reassuring to find that these $\tc$ slopes are positive and that the detailed $\Delta$[X/H] versus $\tc$ relations that we find using our $EW$ measurements are very similar to those obtained by T15. Indeed, the red and blue symbols in their Figure~1 look remarkably similar to our Figure~\ref{f:tc_teske}'s panels b), c), and d). In all these cases, both T15 and ourselves find C, O, S, and Zn abundance differences (N--S) below zero (these are the four elements with the lowest $\tc$ in Figure~\ref{f:tc_teske}). At the same time, virtually all other species show positive differences in both our calculations and T15's. In addition, the average difference for the most refractory elements available is about +0.1\,dex, which combined with the negative differences seen in the volatiles leads to the more steep $\tc$ slopes.

The $\Delta$[X/H](N--S) versus $\tc$ correlation that we have found (our Figure~\ref{f:abNS}) was already seen in the \cite{teske15} data, but with more element-to-element scatter. We believe that part of this scatter is due to the less precise parameters within the ``Alt Params'' sets, but also due to the lower quality of their spectra, which translates to less precise $EW$ values and therefore less accurate relative abundances. On the other hand, it is clear, based on our various tests, that the ``Original Params'' set from T15 does not represent correctly the fundamental properties of the XO-2 stars and that the abundances derived using those parameters are unreliable. Nevertheless, their key finding that ``Fe, Si, and potentially Ni are consistently enhanced in XO-2N'' is fully supported by our study.

\bigskip

The tests made in this section highlight that our relative parameters, and very likely their absolute values, are more precise than those previously published for the XO-2 stars. We attribute this mainly to the higher quality of our data, but also to our use of more reliable effective temperature and surface gravity indicators, even when they lead to inconsistencies in iron-line-only types of analyses. One must realize and accept the fact that those inconsistencies are most likely related to our limitations in the modeling of stellar atmospheres and spectral line calculations, and that they should not be used as arguments against other techniques of stellar parameter determination, especially when these other techniques have proven to be less affected by modeling errors.

\section{PLANET SIGNATURES} \label{s:signatures}

\subsection{Metal depletion from planet formation}

According to \cite{melendez09:twins} and \cite{ramirez11}, planet formation results in a small deficit of metals in the host star's convective envelope relative to its initial composition (and also relative to the composition of the star's radiative zone). The formation of a planet slightly decreases the overall metallicity of the remaining gas in the proto-stellar nebula. If that gas is then accreted onto the host star, its surface metallicity is lowered. Specifically, these studies suggest that when giant planets form, the overall metallicity is changed, while terrestrial planets have a more important effect on the abundances of refractory elements compared to volatiles.

As mentioned in Section~\ref{s:xo-2}, a $0.6\,M_\mathrm{Jup}$ planet has been detected by transit and RV data around XO-2N while two planets with minimum masses $0.26\,M_\mathrm{Jup}$ and $1.4\,M_\mathrm{Jup}$ have been inferred from the RV data of XO-2S. In addition, a long-term RV variability around XO-2N could be interpreted as an additional $1.8\,M_\mathrm{Jup}$ planet, but its strong correlation with stellar activity indicators suggests that it is more likely due to a magnetic cycle instead.

Figure~\ref{f:abNS} shows that XO-2S is metal-poor relative to XO-2N. Simply put, one could say that this result implies that more disk material was used to make planets in XO-2S. This would be compatible with the known planet population of the system if the long-term RV variability of XO-2N is due to magnetic activity and not another giant planet, which appears to be the most natural explanation. Hereafter we assume that the latter is true. From now on, we will also make the reasonable assumption that XO-2 was formed from a single, chemically homogeneous molecular cloud, which fragmented to form two proto-stellar nebulae, each with its own proto-planetary disk.

The very strong correlation between abundance difference and $\tc$ shown in Figure~\ref{f:abNS} complicates the simple explanation given above. Volatiles are just slightly deficient by about 0.015\,dex while refractories are lower by about 0.090\,dex in XO-2S compared to XO-2N. If metals were taken from XO-2S to make more planets, the effect is observed mainly in the refractories. Thus, based on our previous works, the planet population that explains this result cannot be the one detected by the transit and RV data. The planets detected by these traditional techniques have masses that are all greater than those expected for rocky objects. In fact, they are all more massive than Saturn.

Even though the XO-2 planets likely have massive rocky, refractory-rich cores, they are also expected to be surrounded by large volatile rich envelopes. Based on the 16\,Cygni experiment, we expect the formation of these planets to produce mainly a constant offset in the abundance of elements with zero or just a very weak $\tc$ correlation, as in \cite{tucci-maia14}. Thus, only the $\simeq0.015$\,dex offset could be explained by the formation of these planets, and not the $\tc$ trend, or at least not most of it.

Before discussing in more detail the effects described above, we must keep in mind that when interpreting chemical abundance anomalies we are using measurements of photospheric abundances, which are thought to be representative of the composition of the stars' convective envelopes only. If the chemical signatures are imprinted when the stars' radiative zones have fully developed, i.e., after the first few million years since the stars' birth, the amplitudes of these signatures are inversely proportional to the mass of the stars' convective envelopes. In other words, chemical anomalies due to accretion of dust-cleansed or metal-depleted gas will be diluted in stars with massive convective envelopes while they will be easily imprinted in stars with thin convective envelopes. The XO-2 stars are somewhat cooler than the Sun, with $\teff\simeq5500$\,K, but also significantly more metal-rich, at $\feh\simeq+0.4$. For this combination of atmospheric parameters, which imply masses very close to $1\,M_\odot$, the evolutionary tracks by \cite{siess00}, interpolated to the metallicity of the system and using the solar abundances by \cite{asplund09:review} as reference for the scaling, predict a convective envelope mass of $\simeq0.03\,M_\odot$.

\subsection{Metallicity offset: gas giant planets}

Ignoring momentarily the additional depletion of refractories and concentrating only on the 0.015\,dex overall offset of metallicity between XO-2N and XO-2S, we can estimate the gas giant planet mass {\it difference} required to produce it. Note that we refer to a mass difference because both stars are known to host planets and thus metals have been taken away from each of them, but more so for the one that formed more giant planets, i.e., XO-2S. Scaling to the metallicity of the XO-2 stars ($\feh\simeq0.4$), using the \cite{asplund09:review} solar abundances as reference ($(Z/X)_\odot=0.0134$, which leads to $(Z/X)_\mathrm{CZ}=0.034$ for XO-2), and adopting a convective envelope mass $M_\mathrm{CZ}=0.03$\,$M_\odot$ for the XO-2 stars, we can calculate the planet mass difference, provided that we can estimate the metallicities of the known planets, $(Z/X)_p$. As discussed in \citet[][their Section~5.4]{ramirez11}, $(Z/X)_p$ is in the range from 0.04 to 0.12 for Jupiter. As a first order approximation, here we adopt $(Z/X)_p=0.1$ for all XO-2 planets. Under these assumptions, we find that the mass difference that explains a 0.015\,dex offset in [X/H] is about $0.57\,M_\mathrm{Jup}$ (we used the correct form of Equation~1 in \citealt{ramirez11}\footnote{The correct version of that formula has ($M_\mathrm{CZ}+M_p$) instead of only $M_\mathrm{CZ}$ in the denominator. \cite{ramirez11} formula is strictly valid only when $M_p/M_\mathrm{CZ}<<1$.} to figure out this number, $M_p$, which corresponds to the planet mass difference in this case). This value is well below the observed mass difference of the planets detected by transit and RV data, which is {\it at least} 1\,$M_\mathrm{Jup}$ (the latter value is the minimum difference between the total mass of planets known around XO-2S, $1.4\,M_\mathrm{Jup}+0.26\,M_\mathrm{Jup}$, and XO-2N, $0.6\,M_\mathrm{Jup}$). However, if we assume that $(Z/X)_p\simeq0.07$, we can get the mass difference closer to the minimum expected value.

The relative metal-depletion effect described above includes the effect of dilution in the stars' convective envelopes. When we employed $0.03\,M_\odot$ for the mass of these envelopes, we assumed that the planets were all formed after the stars had fully developed their radiative zones and had already thin convective envelopes. However, we know that planets, the gas giants in particular, form within the first few million years from star birth, i.e., while the stars' convective envelopes are more massive than seen during most of their main-sequence lifetimes. Another way of increasing the mass difference computed before to about 1\,$M_\mathrm{Jup}$ given the available data is by assuming that the planets formed when the stars' convective envelopes were about $0.05\,M_\odot$, roughly two-thirds of the way between the time the star was fully convective to the moment its convective envelope reached its final main-sequence mass of $0.03\,M_\odot$. Given all these important considerations, the scenario described here to explain the 0.015\,dex offset of volatiles and refractories for XO-2 is plausible.\footnote{The convective envelope masses that we are assuming are based on our estimates for the total masses of the XO-2 stars, which are both very close to $1\,M_\odot$. When estimating the masses of these stars, we employed our photometrically-derived $\teff$ values, which are about 140\,K warmer than those measured by D15. If we use their atmospheric parameters instead, we would derive masses which are about 5\,\%\ lower, for which the convective envelope masses are about $0.10\,M_\odot$. This would make it easier for us to explain the planet mass difference discussed here. However, we doubt that our stellar parameters are less accurate than those given by D15.}

\subsection{$\tc$ dependence: terrestrial planets}

The rocky bodies of the solar system (meteorites, asteroids, terrestrial planets) are significantly enhanced in refractory elements \cite[e.g.,][]{mcdonough95,alexander01,zuber11}. Interestingly, the anomalous chemical composition of the Sun relative to solar twins mirrors the abundance pattern of meteoritic-like and Earth-like composition rocks. In addition, outside of the solar system, evidence of material with rocky composition is available from observations of white dwarfs polluted by the recent accretion of planetesimals \cite[see, e.g.,][and references therein]{xu14}.

In order to explain the upward $\tc$ trend seen in Figure~\ref{f:abNS}, a depletion of or pollution by rocky, refractory-rich material is required. If indeed related to planetary material, the planets involved would have to be of the terrestrial class. To estimate the amount of rocky material needed to explain the abundance differences seen in Figure~\ref{f:abNS}, we employed a depletion model that follows the scheme of \citet[][see also \citealt{melendez12} and \citealt{ramirez14:bst}]{chambers10}. We calculated the change in photospheric chemical composition that a star experiences due to the formation of rocks for a given mixture of Earth-like and meteoritic-like material, finding the best combination by comparing the results of this calculation to the observations.\footnote{Following \cite{chambers10}, for our calculations we adopted a mix of the chemical composition of the Earth \citep{waenke88} and of CM chondritic meteorites \citep{wasson88}. The rocky material was mixed into a convective envelope mass of $0.03\,M_\odot$, with photospheric abundances obtained by scaling the solar abundances from \cite{asplund09:review}. For each chemical element we computed the change in abundance as 
$\log[(M_\mathrm{CZ}+M_\mathrm{rock,Earth}+M_\mathrm{rock,CM})/M_\mathrm{CZ}]$, where $M_\mathrm{CZ}$ is the mass of the convective zone, $M_\mathrm{rock,Earth}$ is the mass with Earth-like composition, and $M_\mathrm{rock,CM}$ is the mass with CM chondritic composition. We changed the Earth masses of both $M_\mathrm{rock,Earth}$ and $M_\mathrm{rock,CM}$ until a satisfactory fit was found. Using only one type of material does not provide a good fit to both volatile and refractory elements; both Earth-like and meteoritic-like composition are needed to match the observations.} We find that about $20\,M_\oplus$, where $M_\oplus$ is the mass of Earth, of an equal mixture of Earth-like material and meteoritic-like material is necessary to explain the data. This is a significantly larger amount of rocky material than the one required to explain the peculiar solar abundances ($\simeq4\,M_\oplus$), for example \citep{melendez09:twins,chambers10}. Nevertheless, this is naturally explained by the higher metallicity and the larger convective envelope mass of the XO-2 stars.

Two different scenarios could explain the $\tc$ correlation of the abundance differences seen between the XO-2 stars. As in \cite{melendez09:twins}, we could interpret this abundance difference as a signature of the formation of $20\,M_\oplus$ of rocky material around XO-2S, the refractory-poor component. This interpretation does not necessarily imply that no rocky planets formed around XO-2N. However, if rocky planets did form around XO-2N, then the total mass of these planets would have to be lower by about $20\,M_\oplus$ than that of all rocky planets formed around XO-2S. In other words, more refractory-element mass needs to be taken away from XO-2S in order for this scenario to be qualitatively and quantitatively consistent with Figure~\ref{f:abNS}. Within this framework, we would need to explain why XO-2N formed no little planets, or fewer little planets than XO-2S. In any case, none of those planets have yet been detected.

The second explanation that could be provided to account for the $\tc$ correlation is that previously-formed planetary material around XO-2N ended-up being accreted by the star at later times (not necessarily during the formation of star and planets). The amount of this material would have had to be also $20\,M_\oplus$, but if this also happened in XO-2S, then the amount of rocky material accreted by XO-2N would have had to be $20\,M_\oplus$ {\it larger} than that for XO-2S. This idea was already proposed by D15 to explain the iron abundance difference that they found. In this case, one would need to explain why this late accretion of rocks happened only in one of the stars, or more so around one of the stars, namely XO-2N. Perhaps the fact that XO-2N has a gas giant planet in a very close orbit could be related to this phenomenon, since a migrating giant planet can help dragging super-Earths towards the host star. The transiting planet around XO-2N is indeed very close to its parent star, but note also that the two giant planets orbiting XO-2S are relatively close to their star, and have certainly migrated to their present-day orbits.

\subsection{Scenarios not involving planets}

A few studies have contradicted our previous observational results \citep{schuler11} or have provided alternative explanations for the abundance patterns that we have discovered \cite[see, e.g.,][]{gonzalez-hernandez10,gonzalez-hernandez13,onehag11,adibekyan14}. 
Particularly relevant to this paper are two alternative interpretations, 
both unrelated to exoplanets. We address them in turn.

In \cite{onehag11}, it is suggested that the presence of hot stars in the immediate vicinity of the Sun and a solar twin in the M67 open cluster could have affected the final composition of these stars by radiative dust cleansing. While we cannot disprove this idea for the Sun and the solar twin in M67, it appears to be an unlikely scenario for XO-2. Even though the XO-2 stars are widely separated (about 4\,500\,AU), a massive hot star would have had to be much closer to only one of the two stars, XO-2S to be more specific, in order to radiatively eject dust from it while keeping the other star nearly untouched. Dust cleansing by a nearby hot star cannot easily explain refractory-element abundance differences for stars in binary systems and it probably does not work for the case of 16\,Cygni either.

In the work by \cite{adibekyan14} it is suggested that the birthplace of stars in the Galaxy may explain abundance peculiarities that correlate with the elements' condensation temperature. They find that local old stars that have migrated from the inner disk have low refractory-to-volatile ratio. This is definitely not the reason behind our XO-2 result, where one of the components has a low refractory-to-volatile ratio relative to the other one. Under the very reasonable assumption that the XO-2 stars were born together and from the same gas cloud, they share a birthplace and are coeval, yet they exhibit an abundance anomaly of the same amplitude and of very similar nature as that seen in the Sun and a few other solar twins. This experiment as well as our previous work on 16\,Cygni demonstrate that stellar age or birthplace cannot be the only parameters determining the detailed chemical composition of stars and in particular the small anomalies that are detected only with $\simeq0.01$\,dex precision abundances.

\subsection{Other binaries}

Binary systems like 16\,Cygni and XO-2 are particularly interesting in the context of measuring highly-precise relative abundances because each pair consists of two stars which are very similar to each other. Another pair discussed in the literature in the context of planet signatures is HD20782/81, which was analyzed by \cite{mack14}. Planets have been detected around both components in this binary. Contrary to our analyses of 16\,Cygni and XO-2, \citeauthor{mack14} found no difference in the chemical compositions of HD20782 and HD20781. We note, however, that the error bars of their relative abundances are relatively large (average $\Delta\mathrm{[X/H]}=0.04\pm0.07$), most likely owing to the fact that the stars in this system are not twins of each other. Indeed, the $\teff$ difference among them is greater than 400\,K while in 16\,Cygni and XO-2 that difference is only about 60\,K.

In another recent study, our team has investigated the chemical composition of the HAT-P-1 system \citep{liu14}. There, the secondary star hosts a transiting planet of $\simeq0.5\,M_\mathrm{Jup}$, yet the two stars have identical composition at the 0.01\,dex level. One might be tempted to use the HAT-P-1 result by \citeauthor{liu14}\ as evidence against the hypothesis that gas giant planet formation imprints a signature in the chemical composition of the host star. However, we must note that as part of the hypothesis it is required that the stars' convective envelopes are more massive than their final values during the formation of gas giant planets. Since gas giants form quickly, the dilution of their chemical signature by the more massive convective envelopes is facilitated. Thus, it could be that the planet host star in HAT-P-1 had a very massive convective envelope at the time its planet formed. This is not in contradiction with the 16\,Cygni and XO-2 interpretations. In fact, it results naturally from the necessity to either have unusually long-lived disks or from the adoption of episodic accretion models of star formation, in which the specific episodic accretion history of each star determines the time evolution of the convective envelope masses \cite[see our extended discussion of these effects in ][Section~5.4]{ramirez11}.

\subsection{Impact on Galactic Chemical Evolution}

Independently of their true nature, the elemental abundance differences detected between stars in binary systems like 16\,Cygni and XO-2 contradict one of the most important assumptions made in observational Galactic chemical evolution (GCE) studies, namely that the present-day composition of a star reflects that of their parent gas cloud. Certainly, the magnitude of this effect is not so large to introduce significant scatter into traditional GCE-type plots of [X/Fe] versus [Fe/H] such as those presented in the classic GCE paper by \cite{edvardsson93}. However, if 0.01\,dex abundance analyses are done for large samples of stellar twins, this process will become evident and must be taken into account. In particular, these effects may have a non-negligible impact on ongoing and future surveys designed to exploit ``chemical tagging,'' i.e., the association of groups of field stars according to their common composition, which would suggest a common site of formation \citep{freeman02}, as a tool for reconstructing the history of our Galaxy.

\section{FINAL REMARKS}

Stars in multiple systems are expected to have the same surface composition in the absence of accretion effects after the stars have developed thin convective envelopes. Thus, they remove the need to correct for Galactic chemical evolution effects which complicate the study of field stars. If, in addition, the two stars in a given binary are very similar to each other, the complications of standard chemical analysis become irrelevant if a careful strict differential approach is employed. Our careful spectroscopic analysis has revealed a clear difference in the chemical composition of the XO-2 binary system stars, for which a number of possible scenarios involving planets are proposed. 

The XO-2 experiment carried out in this paper provides reliable observational information to further investigate our hypotheses of chemical signatures due to planet formation as well as other alternative explanations. Naturally, high-quality observations of a larger number of twin-star binary systems, both for high-precision chemical analysis and exoplanet search, are highly encouraged to continue this research. 
At the same time, however, we should acknowledge the need to develop a population synthesis modeling approach to explain all the available data in a more robust manner. It should be realized that observational results that characterize a single system cannot be used to fully support or entirely reject our hypotheses. It is becoming increasingly clear that the interpretation of small chemical abundance anomalies in the context of exoplanets is much more complex than previously thought. Of course, this should come as no surprise. After all, we are using this information as indirect, left-over evidence of the very complex processes of star and planet formation.

\acknowledgments

We thank Dr.\ Mario Damasso for providing data tables from D15 in electronic format and the anonymous referee for a very constructive report. D.L.L.\ thanks the Robert A.\ Welch Foundation of Houston, Texas for support through grant F-634. J.M.\ would like to acknowledge support from FAPESP (2012/24392- 2). This work was supported in part by the Australian Research Council (grants FL110100012 and DP120100991).

\end{document}